\documentclass{aa}  

\usepackage{graphicx}
\usepackage{txfonts}
\bibpunct{(}{)}{;}{a}{}{,}
\usepackage{amsmath}
\usepackage{color}
\usepackage{bm}
\usepackage{comment}
\newcommand{\Msun}{\rm{M}_{\sun}} 
\newcommand{\MJup}{\rm{M}_{\rm Jup}} 
\newcommand{\Zsun}{\rm{Z}_{\odot}}
\newcommand{\Eqref}[1]{Equation\:(\ref{#1})}
\newcommand{\Figref}[1]{Figure\:\ref{#1}}
\newcommand{\Figsref}[2]{Figures\:\ref{#1}~and~\ref{#2}}

\newcommand{\bl}[1]{\mbox{\boldmath$ #1 $}}
\begin{document}

   \title{Metallicity dependence of dust growth in a protoplanetary disk}

   \author{Ryoki Matsukoba\inst{1, 2}, 
          Eduard I. Vorobyov\inst{3, 4},
          \and
          Takashi Hosokawa\inst{1}
          }

   \institute{
              Department of Physics, Graduate School of Science, Kyoto University,
              Sakyo, Kyoto 606-8502, Japan \\
              \email{r.matsukoba@tap.scphys.kyoto-u.ac.jp}
         \and
             Center for Computational Sciences, University of Tsukuba, Ten-nodai, 1-1-1 Tsukuba, Ibaraki 305-8577, Japan
         \and
             Department of Astrophysics, University of Vienna, 
             T\''{u}rkenschanzstrasse 17, 1180, Vienna, Austria
         \and
             Research Institute of Physics, Southern Federal University, 
             Roston-on-Don 344090, Russia
             }

   \date{}

   \titlerunning{Dust growth in a low-Z disk}
   \authorrunning{R. Matsukoba et al.}

  \abstract
   {}
   {In the context of planet formation, growth from micron-sized grains to kilometer-sized planetesimals is a crucial question. Since the dust growth rate depends on the amount of dust, realizing planet formation scenarios based on dust growth is challenging in environments with low metallicity, i.e. less dust. We investigate dust growth during disk evolution, particularly focusing on the relationship with metallicity.} 
   {We perform two-dimensional thin-disk hydrodynamic simulations to track the disk evolution over 300\:kyr from its formation. The dust motion is solved separately from the gas motion, with its distribution changing due to drag forces from the gas. Dust size growth is also accounted for, with the magnitude of the drag force varying according to the dust size. We employ three models with metallicities of 1.0, 0.1, and 0.01\:$\Zsun$, i.e. dust-to-gas mass ratios of 10$^{-2}$, 10$^{-3}$, and 10$^{-4}$, respectively.} 
   {In the disks with the metallicities $\ge0.1$\:$\Zsun$, the dust radii reach cm sizes, consistent with estimations from the dust growth timescale. Conversely, for the metallicity of 0.01\:$\Zsun$, the maximum dust size is only 10$^{-2}$\:cm, with almost no growth observed across the entire disk scale ($\sim$100\:au). At the metallicities $\ge0.1$\:$\Zsun$, the decoupling between grown dust and gas leads to non-uniform dust-to-gas mass ratios. However, deviations from the canonical value of this ratio have no impact on the gravitational instability of the disk. The formation of dust rings is confirmed in the innermost part of the disk ($\sim$10--30\:au). The dust rings where the dust-to-gas mass ratio is enhanced, and the Stokes number reaches $\sim$0.1, are suitable environments for the streaming instability. We conjecture that planetesimal formation occurs through the streaming instability in these dust rings.} 
   {}


   \maketitle


\section{Introduction}
\label{Sec:Intro}


Recent observations of exoplanets have highlighted the ubiquity of planetary systems in the present-day Universe, revealing that a multitude of stars host their own planets \citep[NASA Exoplanet Archive;][]{Akeson2013}. Although the foundational mechanisms of planet formation remain shrouded in mystery, efforts to unravel them are progressively advancing \citep[see][for a review]{Drazkowska2023}.


One of the main challenges in planet formation is understanding the growth process from micron-sized grains, prevalent in early protoplanetary disks, to kilometer-sized planetesimals, the building blocks of planets. Dust grains within a protoplanetary disk rotate at Keplerian velocities. Meanwhile, gas is supported by gas pressure and rotates at sub-Keplerian velocities. As a result, dust grains experience a headwind, lose angular momentum, and drift radially toward the central star. In the minimum mass solar nebula \citep[MMSN;][]{Hayashi1981}, when the size of dust grains is in the range of a centimeter to a meter, the drift velocity reaches its maximum, causing the dust to fall into the central star before it can grow further. This phenomenon is known as the ``radial drift barrier'' \citep{Whipple1972, Adachi1976, Weidenschilling1977}. Therefore, suppressing dust drift is essential for the formation of planetesimals.


One of the most classical pictures for planetesimal formation is the gravitational instability of a dust layer accumulated in the midplane of the protoplanetary disk \citep{Safronov1972, Goldreich1973}. However, achieving such a dense dust layer is deemed difficult because turbulence generated by vertical shear stirs up the dust layer \citep{Weidenschilling1980}. Instead, some mechanisms have been proposed to induce dust concentration, such as dust-driven instabilities \citep{Ward1976, Takahashi2014, Tominaga2019, Tominaga2021} and dust trapping by substructures in the disk \citep{Haghighipour2003, Johansen2009, Nayakshin2017b, Vorobyov2019}. The pathway to planetesimal formation through streaming instability is actively investigated by many previous studies \citep[e.g.,][]{Youdin2005, JohansenYoudin2007, Youdin2007, Yang2017, Carrera2022}. The streaming instability, a type of dust-driven instability, arises from the relative drift between dust and gas. This mechanism requires dust sizes in the millimeter to centimeter range in order to moderate coupling with gas. If the streaming instability grows and evolves in a non-linear regime, it leads to strong dust clumping. Subsequently, dust-dense regions collapse into planetesimals due to dust self-gravity once the local dust density exceeds the Roche density \citep[e.g.,][]{JohansenOishi2007, Simon2016}.


In the dust-driven instabilities, it is preferred for the dust grains to be of mm to cm in size, as this size range allows them to moderately decouple from the gas motion by aerodynamic drag. For this reason, the distribution of dust size is crucial for considering these instabilities. Furthermore, high dust-to-gas mass ratios, typically greater than that of solar metallicity (0.01), are required to promote instability. Dust size growth depends on the amount of dust in the disk, with greater amounts of dust leading to faster growth rates. Therefore, disks with high initial dust-to-gas mass ratios, indicative of high metallicity, are advantageous for the realization of these instabilities.


The occurrence rate of planets is correlated with the metallicity of their host stars; however, this correlation varies according to the type of planet. The occurrence rates for gas giants and sub-Neptune planets exhibit a positive correlation with the metallicity of their host stars \citep{Santos2004, Petigura2018}. In contrast, the correlation for super-Earth planets shows a weak dependence on metallicity \citep{Wang2015}, or it may even be absent \citep{Mulders2016, Kutra2021}. These variations in metallicity dependence arise from differences in the formation mechanisms of each type of planet and are likely influenced by the environments in which their precursors, the planetesimals, form.


In this study, we use two-dimensional numerical simulations to consistently track the growth of dust grains during the formation and evolution of the protoplanetary disk. We investigate the impact of the amount of dust on the efficiency of dust grain growth and its effect on the spatial distribution of dust grains, using three models with different metallicities (dust-to-gas mass ratios). Our calculations monitor the disk evolution and dust growth over a period of 300\:kyr following the formation of the disk. This long-term simulation is unprecedented and provides valuable insights into dust growth within a disk of low metallicity, where dust growth is slow.


This paper is organized as follows: We describe our simulation method and setup in Section\:\ref{Sec:method}. We present the simulation results and explain the disk evolution and dust growth for each model in Section\:\ref{Sec:result}. We then discuss the potential for planetesimal and planet formation within the disk in Section\:\ref{Sec:planetesimal}. Summary is given in Section\:\ref{Sec:summary}.

\section{Method}
\label{Sec:method}

We simulate dust growth in protoplanetary disks with different metallicities using the Formation and Evolution of Stars and Disks (FEOSAD) code presented in \citet{VorobyovAkimkin2018} and modified to include the backreaction of dust on gas in \citet{VorobyovElbakyan2020}. FEOSAD solves the equations of hydrodynamics in the thin-disk limit for a gas-dust system. The numerical simulations start from the gravitational collapse of a flattened pre-stellar cloud and the protoplanetary disk is formed self-consistently as a result of angular momentum conservation of the contracting cloud. In this work, for the first time, we applied FEOSAD to non-solar metallicity environments and the usual energy balance equation was modified to include thermal processes that may be important at low metallicity as described in \citet{VorobyovMatsukoba2020}. 


The integration of hydrodynamics equations in the polar coordinates ($r$, \:$\phi$) was carried out using a finite-volume method with a time-explicit solution procedure similar in methodology to the ZEUS code \citep{Stone1992}. The advection of gas and dust is treated using the third-order-accurate piecewise-parabolic interpolation scheme of \citet{Colella1984}. The stellar mass grows according to the mass accretion rate through the inner computational boundary and the properties of the protostar are calculated using the stellar evolution tracks obtained with the STELLAR code for the respective metallicity \citep{Yorke2008, Hosokawa2009}. 


In this work, we are focused on gas and dust dynamics at the intermediate and outer disk regions. Therefore, the central 10\:au were cut out and replaced with a sink cell, while the entire computational domain extends to several thousands of astronomical units. The numerical grid contains $1028 \times 1028$ cells, which are logarithmically spaced in the radial direction and linearly in the azimuthal one, providing sub-au resolution up to a distance of $\approx 150$\:au. Below, we provide the basic equations and initial conditions. More details can be found in the aforementioned articles.

\subsection{Basic equations for the gaseous component}

The system of equations for the gaseous component consists of the continuity equation, equations describing the gas dynamics, and the energy balance equation. The dynamics of gas is governed by the stellar and disk gravity, turbulent viscosity, and friction between gas and dust. The pertinent equations in the thin-disk limit are as follows.
\begin{equation}
\label{eq:cont}
\frac{{\partial \Sigma_{\rm g} }}{{\partial t}}   + \nabla_{\rm p}  \cdot 
\left( \Sigma_{\rm g} {\bl v}_{\rm p} \right) = 0,  
\end{equation}
\begin{eqnarray}
\label{eq:mom}
\frac{\partial}{\partial t} \left( \Sigma_{\rm g} {\bl v}_{\rm p} \right) +  \left[\nabla \cdot \left( \Sigma_{\rm
g} {\bl v}_{\rm p} \otimes {\bl v}_{\rm p} \right)\right]_{\rm p} & =&   - \nabla_{\rm p} {\cal P}  + \Sigma_{\rm g} \, {\bl g}_{\rm p} + \nonumber
\\ 
&+& \left(\nabla \cdot \mathbf{\Pi}\right)_{\rm p}  - \Sigma_{\rm d,gr} {\bl f}_{\rm p},
\end{eqnarray}
where the planar components ($r$,\:$\phi$) are denoted by the subscript ${\rm p}$, $\Sigma_{\rm g}$ is the gas surface density, ${\bl v}_{\rm p}=v_r\hat{{\bl r}}+v_\phi \hat{{\bl \phi}}$ is the gas velocity in the disk plane, $\cal{P}$ is the pressure, integrated in the vertical direction using the ideal equation of state ${\cal P}=(\gamma-1) e$ with the internal energy $e$, and ${\bl f}_{\rm p}$ is the drag force per unit mass between gas and dust. The gravitational acceleration in the disk plane ${\bl g}_{\rm p}$ takes into account gas and dust self-gravity in the disk and the gravity of the central star when it is formed. The gas and dust disk self-gravity is calculated using the convolution method described in detail in \citet{Binney1987}. To compute the viscous stress tensor ${\bl \Pi}$, we parameterise the kinematic viscosity using the usual $\alpha$-parameter approach with the value of $\alpha$ set equal to $10^{-4}$ throughout the disk.


The internal energy balance equation is written as
\begin{equation}
\frac{\partial e}{\partial t} +\nabla_{\rm p} \cdot \left( e {\bl v}_{\rm p} \right) = -{\cal P} 
\left(\nabla_{\rm p} \cdot {\bl v}_{\rm p}\right) -\Lambda  + 
\left(\nabla {\bl v}\right)_{\rm pp^\prime}:\Pi_{\rm pp^\prime}, 
\label{eq:energ}
\end{equation}
where $\Lambda$ encompasses the cooling/heating processes that are pertinent to protoplanetary disks at solar and lower metallicities. In particular, we considered the following processes: the continuum emissions of gas and dust, molecular line emissions of H$_2$ and HD, fine-structure line emissions of O\:{\sc I} ($63\:\mu{\rm m}$) and C\:{\sc II} ($158\:\mu{\rm m}$), and chemical cooling/heating associated with H ionization/recombination and  H$_2$ dissociation/formation. In addition, we solved the non-equilibrium chemical network for eight species, H, H$_2$, H$^+$, H$^-$, D, HD, D$^+$, and e$^-$, with 27 reactions. The evolution of the chemical components affects the cooling rates due to molecular line emissions and chemical cooling. The dust temperature is calculated from the energy balance on dust grains due to the thermal emission, absorption, and collision with gas. This approach permits decoupling of the gas and dust temperatures in the low-density or high-temperature regime or at low metallicity. More details on the thermal scheme applied in this work can be found in \citet{VorobyovMatsukoba2020}.

\subsection{Basic equations for the dust component}

The dust component in our model is divided into two populations: (i) small dust, which are grains with a size\footnote{By the size of dust grains we understand its radius.} between $a_{\rm min}=5\times 10^{-3} \ \mu \rm m$  and $a_{*} = 1 \ \mu \rm m$ and (ii) grown dust ranging in size from $a_{*}$ to a maximum value $a_{\rm max}$, which is variable in space and time. Initially, all dust in a collapsing pre-stellar cloud is in the form of small dust grains. Small dust  can grow and turn into grown dust as the disk forms and evolves. Dust in both populations is distributed over size according to a simple power law: 
\begin{equation}
N(a) = C \cdot a^{ - {\rm q}},
\label{eq:dustdistlaw}
\end{equation} 
where $N(a)$ is the number of dust particles per unit dust size, $C$ is a normalization constant, and ${\rm q} = 3.5$. We note that the power index $\rm q$ is kept constant during the considered disk evolution period for simplicity.


We solve the continuity equations separately for the grown and small dust ensembles. However, the momentum equation is solved only for the grown dust, because small dust is assumed to be dynamically linked to the gas. The system of hydrodynamics equations for the two-population dust ensemble in the zero-pressure limit is written as:
\begin{equation}
\label{contDsmall}
\frac{{\partial \Sigma_{\rm d,sm} }}{{\partial t}}  + \nabla_{\rm p}  \cdot 
\left( \Sigma_{\rm d,sm} {\bl v}_{\rm p} \right) = - S(a_{\rm max}),  
\end{equation}
\begin{equation}
\label{contDlarge}
\frac{{\partial \Sigma_{\rm d,gr} }}
{{\partial t}}  + \nabla_{\rm p}  \cdot 
\left( \Sigma_{\rm d,gr} {\bl u}_{\rm p} \right) =  
S(a_{\rm max}),
\end{equation}
\begin{eqnarray}
\label{eq:momDlarge}
\frac{\partial}{\partial t} \left( \Sigma_{\rm d,gr} {\bl u}_{\rm p} \right) +  \left[\nabla \cdot \left( \Sigma_{\rm d,gr} {\bl u}_{\rm p} \otimes {\bl u}_{\rm p} \right)\right]_{\rm p}  &=&   \Sigma_{\rm d,gr} \, {\bl g}_{\rm p} + \nonumber \\
 + \Sigma_{\rm d,gr} \, {\bl f}_{\rm p} + S(a_{\rm max}) {\bl v}_{\rm p},
\end{eqnarray}
where $\Sigma_{\rm d,sm}$ and $\Sigma_{\rm d,gr}$ are the surface densities of small and grown dust, respectively, and ${\bl u_{\rm p}}$ are the planar components of the grown dust velocity.


The grown dust dynamics is sensitive to the properties of surrounding gas. The drag force (per unit mass) links dust with gas and can be written according to \citet{Weidenschilling1977} as:
\begin{equation}
    {{\bl f}_{\rm p}} = \dfrac{1} {2 m_{\rm d}} C_{\rm D} \, \sigma \rho_{\rm g} ({{\bl v_{\rm p}}} - {{\bl u_{\rm p}}}) |{{\bl v_{\rm p}}} - {{\bl u_{\rm p}}}|,
\label{eq:friction}
\end{equation}
where $\sigma$ is the dust grain cross section, $\rho_{\rm g}$ the volume density of gas, $m_{\rm d}$ the mass of a dust grain, and $C_{\rm D}$ the dimensionless friction parameter. The functional expression for the latter is taken from \citet{Henderson1976} and is described in detail in   \citet{Vorobyov2023}. The use of the Henderson friction coefficient allows us to treat the drag force avoiding discontinuities in two different regimes: the Epstein regime, and Stokes linear and non-linear regimes. The advantages of the Henderson friction coefficient over the more common drag coefficient proposed by \citet{Weidenschilling1977} are discussed in \citet{Stoyanovskaya2020}. To account for the back-reaction of grown dust on dust, the term $\Sigma_{\rm d,gr} f_{\rm p}$ is symmetrically included in both the gas and dust momentum equations. We use the maximum size of dust grains $a_{\rm max}$ when calculating the value of $m_{\rm d}$ in Equation~(\ref{eq:friction}). The friction force $f_{\rm p}$ thus derived would describe the dynamics of the main dust mass carriers, since large grains near $a_{\rm max}$ mostly determine the value of $m_{\rm d}$ for the chosen value of ${\rm q}=3.5$. 


The term $S(a_{\rm max})$ that enters the equations for the dust component is the conversion rate between small and grown dust populations. We assumed that the distribution of dust particles over size follows the form given by Equation\:\eqref{eq:dustdistlaw} for both small and grown populations. Furthermore, the distribution is assumed to be continuous at $a_{*}$. Our scheme is constructed so as to preserve continuity at $a_{*}$ by writing the conversion rate of small to grown dust in the following form:
\begin{equation}
    S(a_{\rm max}) = - \dfrac{\Delta \Sigma_{\rm d,sm}} {\Delta t},
\end{equation}
where
\begin{equation}
\label{final}
    \Delta\Sigma_{\mathrm{d,sm}} = \Sigma_{\mathrm{d,sm}}^{n+1}- \Sigma_{\mathrm{d,sm}}^{n} =
    \frac
    {
    \Sigma_{\rm d,gr}^n \int_{a_{\rm min}}^{a_*} a^{3-\mathrm{q}}da - 
    \Sigma_{\rm d,sm}^n \int_{a_*}^{a_{\mathrm{max}}^{\rm n+1}} a^{3-\mathrm{q}}da
    }
    {
    \int_{a_{\rm min}}^{a_{\mathrm{max}}^{n+1}} a^{3-\mathrm{q}}da
    },
\end{equation}
where indices $n$ and $n+1$ denote the current and next hydrodynamic steps of integration, respectively, and $\Delta t$ is the hydrodynamic time step. The adopted scheme effectively assumes that dust growth smooths out any discontinuity in the dust size distribution at $a_\ast$ that may appear due to differential drift of small and grown dust populations. A more detailed description of the scheme is presented in~\citet{Molyarova2021} and \citet{VorobyovSkliarevskii2022}. 


The conversion rate $S(a_{\rm max})$ depends only on the local maximal size of dust $a_{\rm max}$, since the values of $a_{\rm min}$ and $a_\ast$ are fixed in our model. At the beginning of the cloud core collapse, all grains are in the form of small dust, namely, $a_{\rm max} = 1.0$\:$\mu$m. During the disk formation and evolution epoch the maximal size of dust particles usually increases. The change in $a_{\rm max}$ within a particular numerical cell in our model occurs due to collisional growth or via advection of dust through the cell. The equation describing the dynamical evolution of $a_{\rm max}$ is as follows:
\begin{equation}
\frac{\partial a_{\rm max}}{\partial t} + ({\bl u}_{p} \cdot \nabla_p ) a_{\rm max} = \cal{D},
\label{eq:dustA}
\end{equation}
where the rate of dust growth due to collisions and coagulation is computed in the monodisperse approximation \citep{Birnstiel2012}
\begin{equation}
\cal{D} = \frac{\rho_{\rm d} \mathit{u}_{\rm rel}}{\rho_{\rm s}}.
\end{equation}
This rate includes the total volume density of dust $\rho_{\rm d}$, the dust material  density $\rho_{\rm s} = 3.0$\:g\:cm$^{-3}$, and the relative velocity of particle-to-particle collisions defined as $\mathit{u}_{\rm rel} = (\mathit{u}_{\rm th}^2 + \mathit{u}_{\rm turb}^2)^{1/2}$, where $\mathit{u}_{\rm th}$ and $\mathit{u}_{\rm turb}$ account for the Brownian and turbulence-induced local motion, respectively. When calculating the volume density of dust, we take into account dust settling by calculating the effective scale height of grown dust  $H_{\rm d}$ via the corresponding gas scale height $H_{\rm g}$, $\alpha$ parameter, and the Stokes number as
\begin{equation}
    H_{\rm d} = H_{\rm g} \sqrt{\frac{\alpha}{\alpha + \mathrm{St}}}.
    \label{eq:dust-scale-height}
\end{equation}
The Stokes number is defined as:
\begin{equation}
    {\rm St} = \frac{\Omega_{\rm K}\rho_{\rm s}a_{\rm max}}{\rho_{\rm g}c_{\rm s}},
\end{equation}
where $\Omega_{\rm K}=\sqrt{GM_{\ast}/r^{3}}$ represents the Keplerian angular velocity with the central stellar mass $M_{\ast}$, $\rho_{\rm g}$ is the volume density of gas, and $c_{\rm s}$ is the sound speed.


Dust growth in our model is limited by collisional fragmentation and drift.  We note that the drift barrier is accounted for self-consistently via the computation of the grown dust dynamics. The fragmentation barrier is taken into account by calculating the characteristic fragmentation size as \citep{Birnstiel2016}:
\begin{equation}
    a_{\rm frag}=\frac{2\Sigma_{\rm g} \mathit{u}_{\rm frag}^2}{3\pi\rho_{\rm s} \alpha c_{\rm s}^2},
\label{eq:afrag}
\end{equation}
where $\mathit{u}_{\rm frag}$ is the fragmentation velocity, namely, a threshold value of the relative velocity of dust particles at which collisions result in fragmentation rather than coagulation. In the current study, we adopt $\mathit{u}_{\rm frag} = 3$\:m\:s$^{-1}$ \citep{Blum2018}. If $a_{\rm max}$ becomes greater than $a_{\rm frag}$, we halt the growth of dust and set $a_{\rm max} = a_{\rm frag}$.

\begin{table*}
\center
\caption{\label{Tab:1}The initial parameters of the pre-stellar cloud cores}
\begin{tabular}{l c c c c c c c c}
\hline 
\hline 
Metallicity & Dust-to-gas mass ratio & $\Sigma_0$ & $c_{{\rm s},0}$ & $T_0$ & $\rho_0$ & $r_0$ & $\Omega_0$ & $M_{\rm cloud}$ \tabularnewline
$\Zsun$ &  & g\:cm$^{-2}$ & km\:s$^{-1}$ & K & g\:cm$^{-3}$ & au & km\:s$^{-1}$\:pc$^{-1}$ & $\Msun$ \tabularnewline
\hline
1.0   & $10^{-2}$ & $9.0\times10^{-2}$ & 0.19 & 10 & $3.8\times10^{-18}$ & 1600 & 2.1 & 0.88 \tabularnewline
0.1   & $10^{-3}$ & $9.8\times10^{-2}$ & 0.21 & 12 & $3.8\times10^{-18}$ & 1700 & 2.1 & 0.88 \tabularnewline
0.01  & $10^{-4}$ & $0.70$             & 0.33 & 31 & $7.6\times10^{-17}$ & 620  & 9.4 & 0.89 \tabularnewline
\hline
\end{tabular}
\center{ \textbf{Notes.} $\Sigma_0$ is the surface density at the plateau, $c_{{\rm s},0}$ is the initial sound speed, $T_0$ is the initial temperature, $\rho_0$ is the initial mass density, $r_0$ is the plateau radius, $\Omega_0$ is the angular velocity at the plateau, and $M_{\rm cloud}$ is the cloud mass.}
\end{table*}

\subsection{Initial conditions}
\label{Sect:initial}

The initial gas surface density and angular velocity profiles of the cloud core we set are derived from an axisymmetric cloud collapse where the angular momentum remains constant \citep{Basu1997}:
\begin{equation}
\Sigma_{\rm g} = \frac{\Sigma_{0}}{\sqrt{1 + \left( r/r_{0} \right)^2}},
\label{Eq:Sigma0}
\end{equation}
\begin{equation}
\Omega = 2\Omega_{0}\left( \frac{r_{0}}{r} \right)^2 \left[ \sqrt{1 + \left( \frac{r}{r_{0}} \right)^2} - 1 \right].
\label{Eq:Omega0}
\end{equation}
The above profile has a plateau with a uniform surface density extending to the radius $r_{0}$, $\Sigma_{0}$ and $\Omega_{0}$ are the surface density and angular velocity at the plateau, respectively. The plateau radius $r_{0}$ is proportional to the Jeans length
\begin{align}
r_{0} = A \frac{c_{{\rm s},0}}{\sqrt{\pi G \rho_{0}}}, 
\label{Eq:r0}
\end{align}
where $A$ is a constant parameter, $c_{{\rm s},0}$ is the initial sound speed, $G$ is the gravitational constant, and $\rho_{0}$ is the initial mass density. The initial mass and the plateau surface densities are related as $\Sigma_{0} = r_{0}\rho_{0}$. The constant $A$ is a parameter defining the initial density perturbation. We set $A = \sqrt{1.2}$, with which the ratio of the cloud thermal and gravitational energies is 0.8. We also set the ratio of the rotational and gravitational energies as $7\times10^{-3}$ by adjusting the plateau angular velocity $\Omega_{0}$. 


We calculate three models with different metallicities, i.e., different dust-to-gas mass ratios. These models have metallicities of 1.0, 0.1, and 0.01\:$\Zsun$, with corresponding dust-to-gas mass ratios of $10^{-2}$, $10^{-3}$, and $10^{-4}$, respectively. The initial dust-to-gas mass ratio in the gravitationally contracting cloud is spatially uniform. 


To constrain the free parameters of the initial surface density configuration, namely, $c_{{\rm s},0}$ and $\rho_0$, we make use of a one-zone model developed by \cite{Omukai2005}, which derives the relationship between the central density and temperature in clouds of different metallicity. We note that the gas temperature varies with metallicity and models with different metallicities are characterized by distinct initial sound speeds $c_{{\rm s},0}$. To simulate the evolution of the disk formed from a cloud of nearly solar mass, we select densities and temperatures at which the Jeans mass approximates solar mass, based on the derived density-temperature relationship of \citet{Omukai2005}. These selections help determine the initial mass density $\rho_{0}$ and the initial sound speed $c_{{\rm s},0}$. The plateau radius is then calculated from Equation\:(\ref{Eq:r0}) using these initial conditions, and the initial gas surface density $\Sigma_0$ is determined as $\Sigma_{0} = r_{0}\rho_{0}$. Table\:\ref{Tab:1} summarizes the values for the initial pre-stellar cloud conditions in each model. Regarding the dust surface density, we assume that it consists solely of small dust grains, and its density is scaled from the gas surface density in accordance with the dust-to-gas mass ratio.


The initial density is perturbed with an amplitude $A$, leading to gravitational collapse of the pre-stellar cloud. The cloud contracts, while simultaneously spinning up. The protoplanetary disk forms when the in-spiraling material of the contracting cloud hits the centrifugal barrier near the inner computational boundary. This occurs when the centrifugal radius of the infalling gas exceeds the sink cell radius of 10\:au after about twice the free-fall time elapsed from the onset of cloud collapse. Subsequently, the disk gains matter from the infalling parental cloud while losing it via accretion onto a star introduced within the sink cell. We follow the evolution of the disk and the growth of dust for 300\:kyr after the disk formation.

\section{Global disk evolution and dust growth}
\label{Sec:result}

Here, we describe the global evolution of a protoplanetary disk and dust growth within the disk at three different metallicities. We present our simulation results for the cases of solar metallicity and low metallicities in Sections\:\ref{Sec:solar} and \ref{Sec:lowZ}, respectively. Since the results for the solar metallicity case are similar to those of previous studies \citep[e.g.,][]{Vorobyov2018, Vorobyov2019}, we provide only a brief overview of the solar metallicity case.

\subsection{Solar metallicity}
\label{Sec:solar}

\begin{figure}
 \begin{center}
 \begin{tabular}{c} 
  {\includegraphics[width=0.95\columnwidth]{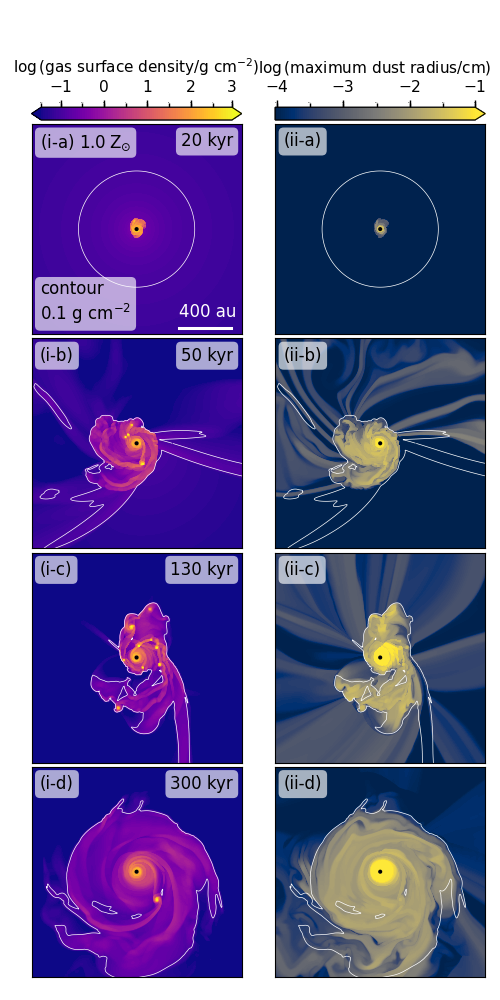}}
 \end{tabular}
 \caption{
 Time evolution of the disk in the case of 1.0\:$\Zsun$. 
 The left and right columns show the 2D distributions of the gas surface density and the maximum dust size at four different epochs, 20, 50, 130, and 300\:kyr after the formation of the disk. In each panel, the white contour line delineates an isosurface density of 0.1\:g\:cm$^{-2}$, indicating the approximate location of the outer edge of the disk.
 }
 \label{Fig:1}
 \end{center}
\end{figure}


Figure\:\ref{Fig:1} shows the evolution of the disk from its formation to the last period at 300\:kyr. The panels in the left column display the spatial distributions of gas surface density, indicating that the disk's size expands over time. The white contour lines in the panels show isosurface density lines of 0.1\:g\:cm$^{-2}$, serving as a rough indicator of the outer edge of the disk. We note that at 20\:kyr (panel i-a), the density at the outer edge exceeds 0.1\:g\:cm$^{-2}$, leading to a disparity between the outer edge and the contour line. The disk extends to approximately 200--300\:au at 50\:kyr, and around 600\:au at 300\:kyr.


Panels i-b, i-c, and i-d show that, at 50, 130, and 300\:kyr, non-axisymmetric structures like spiral arms appear within the disk, and several clumps exist. This suggests that the disk is gravitationally unstable. In particular, disk fragmentation is most intense around 100\:kyr (panel i-c). Thereafter, it gradually calms down, and only one clump remains within the disk at 300\:kyr (panel i-d).


In panels i-b and i-c, there are zonal regions with densities exceeding 0.1\:g\:cm$^{-2}$, which extend beyond the panel boundaries. These regions represent remnants of clumps ejected from the central part. Multiple ejection events are observed between 30 and 210\:kyr in this run. These events, triggered by gravitational interactions and close encounters between clumps, occur more frequently during periods of intense fragmentation. Ejected clumps escape from the stellar system and may serve as a source for freely floating brown dwarfs and even planetary-mass objects \citep[][]{Bate2009, Stamatellos2009, Basu2012, Vorobyov2016}.


The panels in the right column of Figure\:\ref{Fig:1} display the spatial distributions of the maximum dust size. At 20\:kyr (panel ii-a), dust growth is observed within a region inside $\sim$100\:au, corresponding to the disk radius. Subsequently, as the disk radius expands, the area containing large dust grains ($>10^{-3}$\:cm) also increases. At 300\:kyr (panel ii-d), the regions with the large dust grains mainly lie inside the white isodensity lines, indicating that dust growth occurs within the disk. Panels ii-b and ii-c indicate the ejection events, characterized by dust grains ranging from $10^{-3}$ to $10^{-2}$\:cm in size, dispersing radially from the outer edge of the disk.

\begin{figure}
 \begin{center}
 \begin{tabular}{c} 
  {\includegraphics[width=0.94\columnwidth]{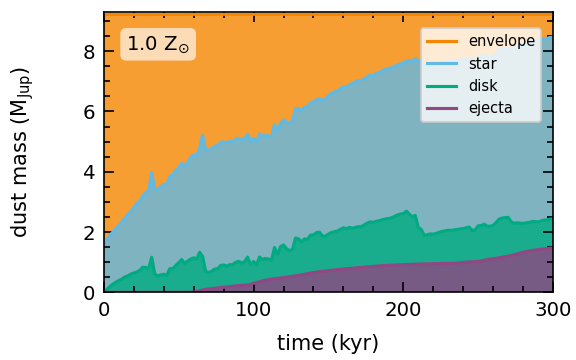}}
 \end{tabular}
 \caption{
 Stacked area chart for the dust mass in the case of 1.0\:$\Zsun$. The dust mass is the sum of the masses of grown and small dust grains. The colors represent four different components: the envelope (orange), central star (blue), disk (green), and ejecta (purple). 
 }
 \label{Fig:2}
 \end{center}
\end{figure}

\Figref{Fig:2} presents a stacked area chart for dust mass and indicates that fifteen percent of the total dust ($\sim$1.5\:$\MJup$) is ejected from the stellar system. Observations at mm wavelengths show that the dust opacity spectral index is around or smaller than unity in the envelopes of Class 0 and I objects \citep[e.g.,][]{Kwon2009, Chiang2012, Miotello2014, Galametz2019}, whereas it is $\sim$1.7 in the interstellar medium \citep{Finkbeiner1999, Li2001}. These observations suggest the presence of grown dust grains larger than 1\:$\mu$m in the envelope at an early stage of star formation. Such dust grains are likely supplied by the outflow from the disk \citep{Wong2016, Tsukamoto2021} rather than being formed in situ \citep{Ormel2009}, due to short timescales of $\sim10^5$\:years. Our results imply that not only the outflow but also the ejection can transport grown dust grains to the envelope.

\begin{figure*}
 \begin{center}
 \begin{tabular}{c} 
  {\includegraphics[width=1.98\columnwidth]{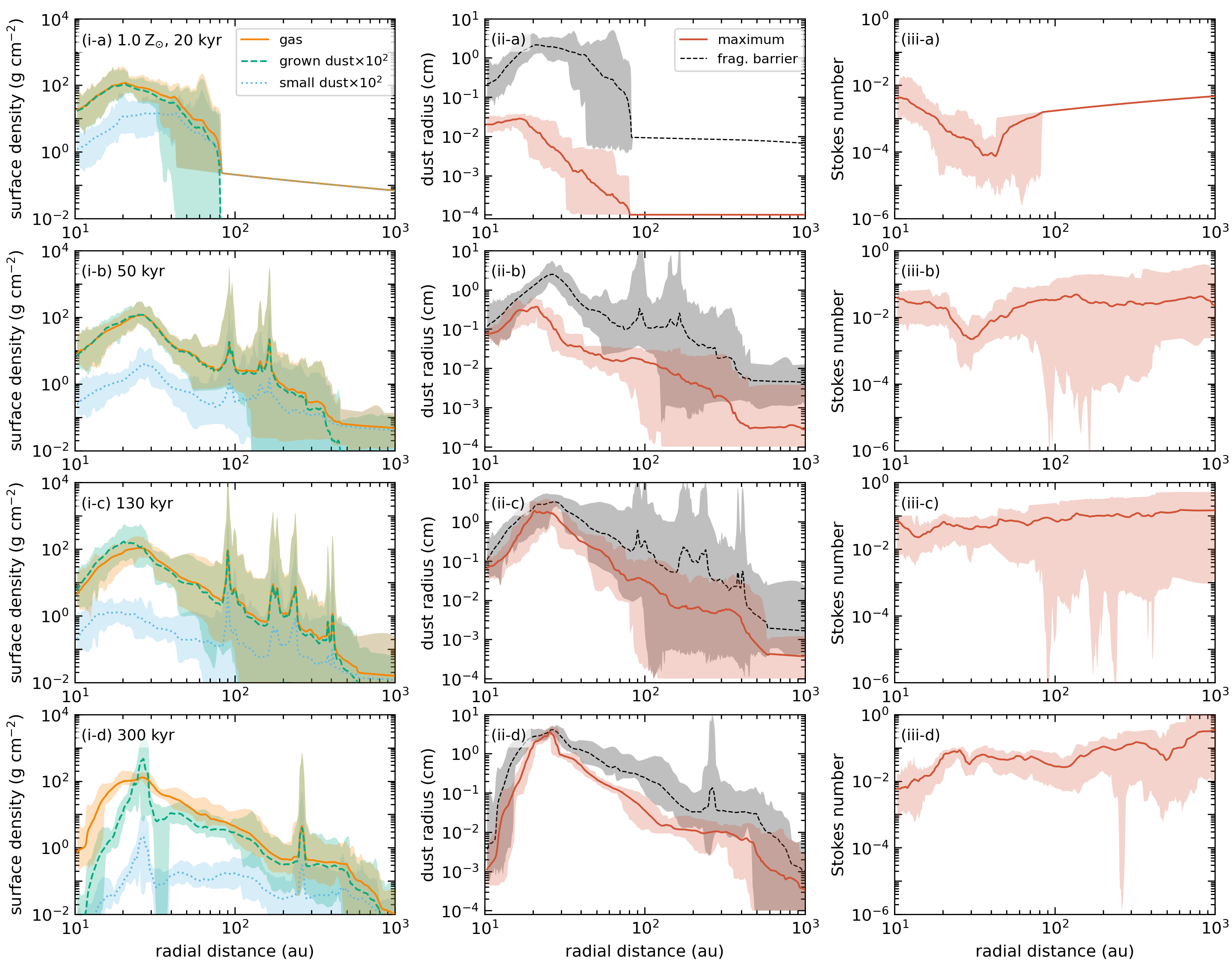}}
 \end{tabular}
 \caption{
 Evolution of the radial distributions of the gas and dust for the case of 1.0\:$\Zsun$. 
 Each column shows the radial profiles of the surface densities (left), the maximum dust size (middle), and the Stokes number (right) at the same epochs as in Figure\:\ref{Fig:1}. 
 The lines provide the azimuthally averaged values, and the shaded layers show the ranges of variation of these values at a given radius, with the upper and lower boundaries corresponding to the maximum and minimum values. 
 In the left panels, the orange, green, and blue colors represent different components: gas, grown dust, and small dust, respectively. 
 The densities of grown and small dust are multiplied by 100 to make comparisons easier. 
 In the middle panels, the lines depict the maximum dust size (red solid line) and the fragmentation barrier given by Equation\:(\ref{eq:afrag}) (black dashed line). 
 }
 \label{Fig:3}
 \end{center}
\end{figure*}


Figure\:\ref{Fig:3} illustrates the radial distributions of various properties of gas and dust grains. In the left column, the surface densities of gas, grown dust, and small dust are displayed, allowing a quantitative comparison of their magnitudes. For example, panel i-a shows a gradual increase in the gas density toward the center from 1000 to 80 au, indicating the presence of the accretion envelope. A significant order-of-magnitude increase in gas density is observed at 80 au, defining the outer boundary of the disk. The transition between the envelope and the disk is characterized by a shift in dominant dust components; the envelope contains primarily small dust, whereas the disk is rich in grown dust grains. Panels i-a and i-b show that, at 20 and 50\:kyr, the gas surface density is approximately equal to 100 times the surface density of grown dust in the inner regions, and 100 times the surface density of small dust in the outer regions. By the epoch of 130\:kyr (panel i-c), a discrepancy emerges between the gas and grown dust distributions within a radius of 100\:au. This difference becomes more prominent at 300\:kyr (panel i-d), where the surface density of the grown dust exhibits a peak at around 30\:au. The dust-to-gas mass ratio exceeds the conventional value of $10^{-2}$ within such a "dust ring". At 300\:kyr, there is a depletion of dust grains due to their radial drift motion, resulting in the total dust mass (summing up the grown and small components) being $0.76\times10^{-2}$ of the gas mass within 1000\:au, despite their comparable masses up to 130\:kyr.


The dust ring is created by accumulation of radially drifting dust at a gas pressure bump. In a disk that is gravitationally unstable, the transport of angular momentum is more effective in the outer regions, where the gravitational instability is most pronounced, and becomes less effective towards the inner regions. As the efficiency of angular momentum transport varies with distance from the center, gas flowing towards the center piles up in an innermost region of the disk, leading to the formation of a pressure bump. This phenomenon, known as the "bottleneck effect," has been extensively studied by \citet{Vorobyov2023}. We note that the dust ring is at a radial position where the efficiency of angular momentum transport (i.e., gravitational and viscous torque) takes a minimum, which depends on the phenomenological $\alpha$ parameter. Additionally, its position may also depend on the sink radius, with a smaller sink radius causing the dust ring to shift inward (see \citealt{Vorobyov2023} for more details).


The middle panels in Figure\:\ref{Fig:3} demonstrate efficient dust growth within the disk, where the maximum dust size is significantly larger than in the surrounding envelope. We find that the dust size grows to mm scale at an early stage, reaching 0.1 to 1.0\:mm within 100\:au at 50\:kyr (panel ii-b). The timescale for the collisional dust growth from micron to mm radii is given by \citep{Birnstiel2016} 
\begin{align}
  t_{\rm growth} &\simeq \left( \frac{\Sigma_{\rm d}}{\Sigma_{\rm g}} \right)^{-1} \frac{1}{\Omega_{\rm K}} \notag \\
  &\sim 20 \left( \frac{\Sigma_{\rm d}/\Sigma_{\rm g}}{10^{-2}} \right)^{-1} \left( \frac{M_{\ast}}{0.5\:\Msun} \right)^{-1/2} 
  \left( \frac{r}{100\:{\rm au}} \right)^{3/2}\:{\rm kyr}.
  \label{Eq:t_growth}
\end{align}
Note that collisions between dust grains of equal size are assumed in the above estimate, as used in our dust growth scheme. 
Our results are consistent with this timescale. After 50\:kyr, the dust size nearly reaches a plateau of growth (panels\:ii-c and ii-d). Except in the innermost regions, the fact that the maximum dust size stays below the fragmentation barrier suggests that the drift barrier (i.e., radial drift) determines the dust size. 


The Stokes number increases with the growth of dust grains. In Figure\:\ref{Fig:3}, panel iii-a shows that in the early stages, the Stokes number varies from $10^{-2}$ to $10^{-4}$ within the disk. As dust grows further, the Stokes number reaches around 0.1 throughout the disk. The lower Stokes number in the early stages is not only due to the small size of dust grains but also to the high density and temperature commonly observed in young disks \citep{Vorobyov2024}. Moreover, the dust ring exhibits a high dust-to-gas mass ratio and a Stokes number exceeding 0.01, indicating the potential occurrence of streaming instability in this region. We will further discuss this possibility later in Section\:\ref{Sec:planetesimal}. 

\begin{figure}
 \begin{center}
 \begin{tabular}{c} 
  {\includegraphics[width=0.94\columnwidth]{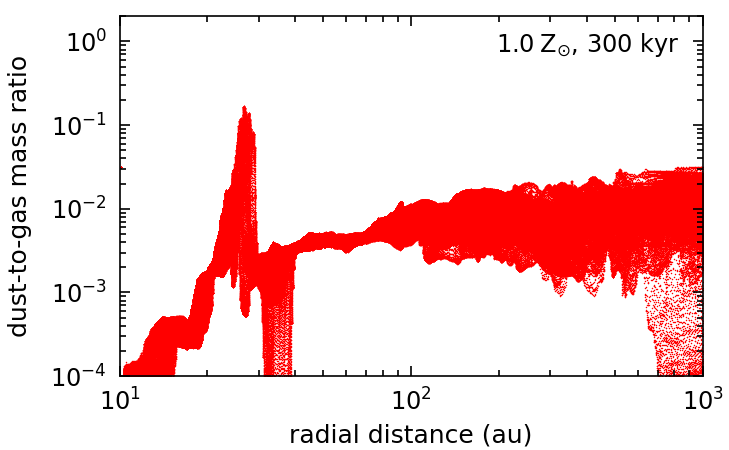}}
 \end{tabular}
 \caption{
 Variation in the dust-to-gas mass ratio from the canonical value in the case of 1.0\:$\Zsun$ at 300\:kyr. The red dots represent the dust-to-gas mass ratios for each grid cell.
 }
 \label{Fig:4}
 \end{center}
\end{figure}

Figure\:\ref{Fig:4} presents the variation of the dust-to-gas mass ratio at the epoch of 300\:kyr. We see that the dust-to-gas mass ratio is as high as $\sim$ 0.1 at the dust ring, whose radius is approximately 30\:au. In the outer part of the ring within $\sim 100$\:au, the dust-to-gas mass ratio is slightly smaller than the initial value 0.01, suggesting that some grains have drifted inward to be trapped by the dust ring. Moving even further out beyond 100\:au, there are large scatters due to the gravitational instability and turbulent motion in that region. However, the scatters are insufficient to alter the thermal conditions of the gas and, consequently, the strength of the disk fragmentation. Throughout the evolution, we observe almost the same level of disk fragmentation as in the case where we ignore the dust growth and drifting motion apart from the gas.

\subsection{Low metallicities}
\label{Sec:lowZ}

We describe our simulation results for the low-metallicity cases with 0.1 and 0.01\:$\Zsun$ in Sections\:\ref{Sec:0.1Zsun} and \ref{Sec:0.01Zsun}, respectively. In particular, we study disk evolution and dust growth at these metallicities, focusing on their differences from the solar-metallicity case.

\subsubsection{One tenth of solar metallicity}
\label{Sec:0.1Zsun}

\begin{figure}
 \begin{center}
 \begin{tabular}{c} 
  {\includegraphics[width=0.95\columnwidth]{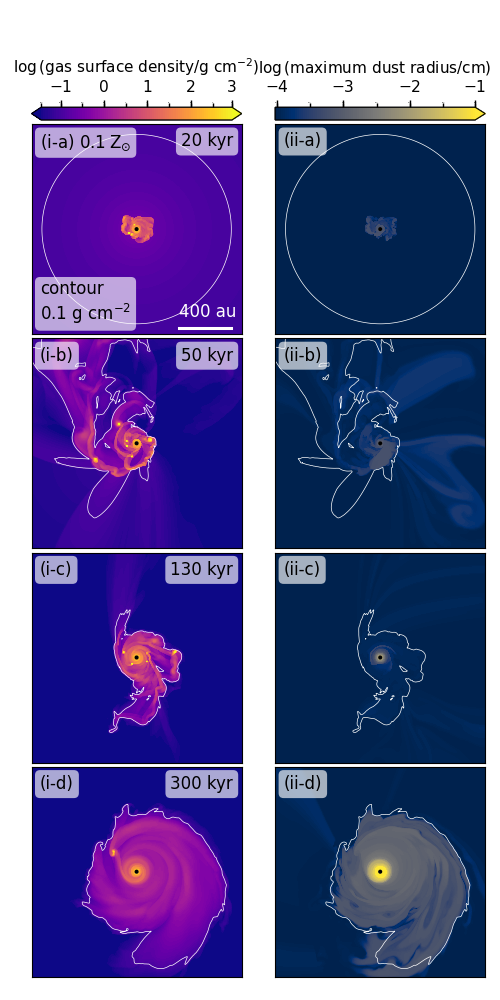}}
 \end{tabular}
 \caption{
 Same as \Figref{Fig:1} but for the case of 0.1\:$\Zsun$
 }
 \label{Fig:5}
 \end{center}
\end{figure}


Figure\:\ref{Fig:5} shows the evolution of the disk at 0.1\:$\Zsun$. It can be observed from the left panels that the outer radius of the disk expands over time, reaching approximately 600\:au by the last epoch at 300\:kyr. Disk fragmentation is evident, particularly at 130\:kyr, where fragmentation is intense with $\sim$ 10 clumps present in the disk (panel i-c). At 300\:kyr, the disk fragmentation diminishes (but gravitational instability persists), with only one clump remaining (panel i-d). The evolution of the disk up to this epoch closely resembles that of the solar-metallicity disk, as shown in Figure\:\ref{Fig:1}. The similarity in the disk evolution at 1.0 and 0.1\:$\Zsun$ has been also reported by previous studies that investigated the metallicity dependence of the disk fragmentation (See \citealt{Tanaka2014} for an analytic study; \citealt{Machida2015, Bate2014, Bate2019, Vorobyov2020, Matsukoba2022} for numerical studies). Our results support their findings.


Our simulation for the case with 0.1\:$\Zsun$ shows two ejection events: one occurring at 20\:kyr (shortly after the epoch of panel i-a) and the other at 70\:kyr. In panel i-b, the regions that extend from the center to the upper left, where the density exceeds 0.1\:g\:cm$^{-2}$, provide evidence of the first ejection event at 20\:kyr. Similarly to the case with solar metallicity, the ejection events are triggered by close encounters between clumps, and a portion of the ejected material escapes from the stellar system. 


In Figure\:\ref{Fig:5}, panels ii-a and ii-b indicate that the largest dust size present in all areas is approximately $10^{-4}$\:cm, showing no indications of dust growth. By 130\:kyr (panel ii-c), the grown dust grains appear only in an innermost part of the disk (around 10--20\:au), although the maximum dust size remains at $10^{-4}$\:cm across the entire disk. The spatial distribution of the maximum dust size dramatically changes at 300\:kyr (panel ii-d), with dust growth evident throughout the disk. In the outer part of the disk, the maximum dust size eventually reaches $\sim10^{-3}$ to $10^{-2}$\:cm. This contrasts to the case with 1.0\:$\Zsun$, where the disk becomes filled with dust grains $\ge10^{-2}$\:cm much earlier. Furthermore, unlike the case with 1.0\:$\Zsun$, the maximum dust size within the envelope remains at $\sim10^{-4}$\:cm at all periods, indicating the absence of grown dust. This is because the maximum dust size within the disk remains small during ejection events at 20\:kyr and 70\:kyr, preventing the transport of large dust grains to the envelope through ejection.

\begin{figure}
 \begin{center}
 \begin{tabular}{c} 
  {\includegraphics[width=0.94\columnwidth]{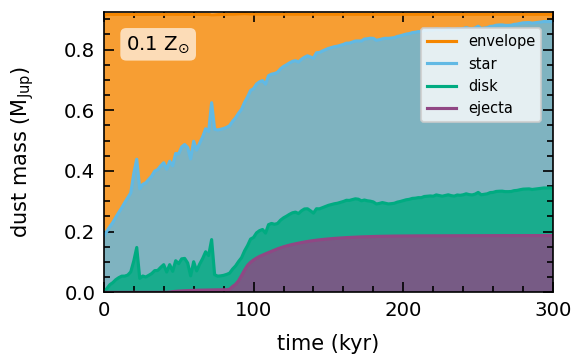}}
 \end{tabular}
 \caption{
 Same as Figure\:\ref{Fig:2} but for the cases of 0.1\:$\Zsun$.
 }
 \label{Fig:6}
 \end{center}
\end{figure}

Figure\:\ref{Fig:6} presents a stacked area chart of the dust mass for the case of 0.1\:$\Zsun$, revealing that dust grains with a mass equivalent to one-fifth of the initial cloud (0.2\:$\MJup$) are ejected from the system. Most of the ejecta are provided by the event at 70\:kyr. As evident from Figure\:\ref{Fig:5}, dust size growth is hardly progressed by the time of the ejection events. Therefore, only small dust grains, not grown dust grains, are scattered into the surrounding envelope by ejection events, unlike in the case of solar metallicity.

\begin{figure*}
 \begin{center}
 \begin{tabular}{c} 
  {\includegraphics[width=1.98\columnwidth]{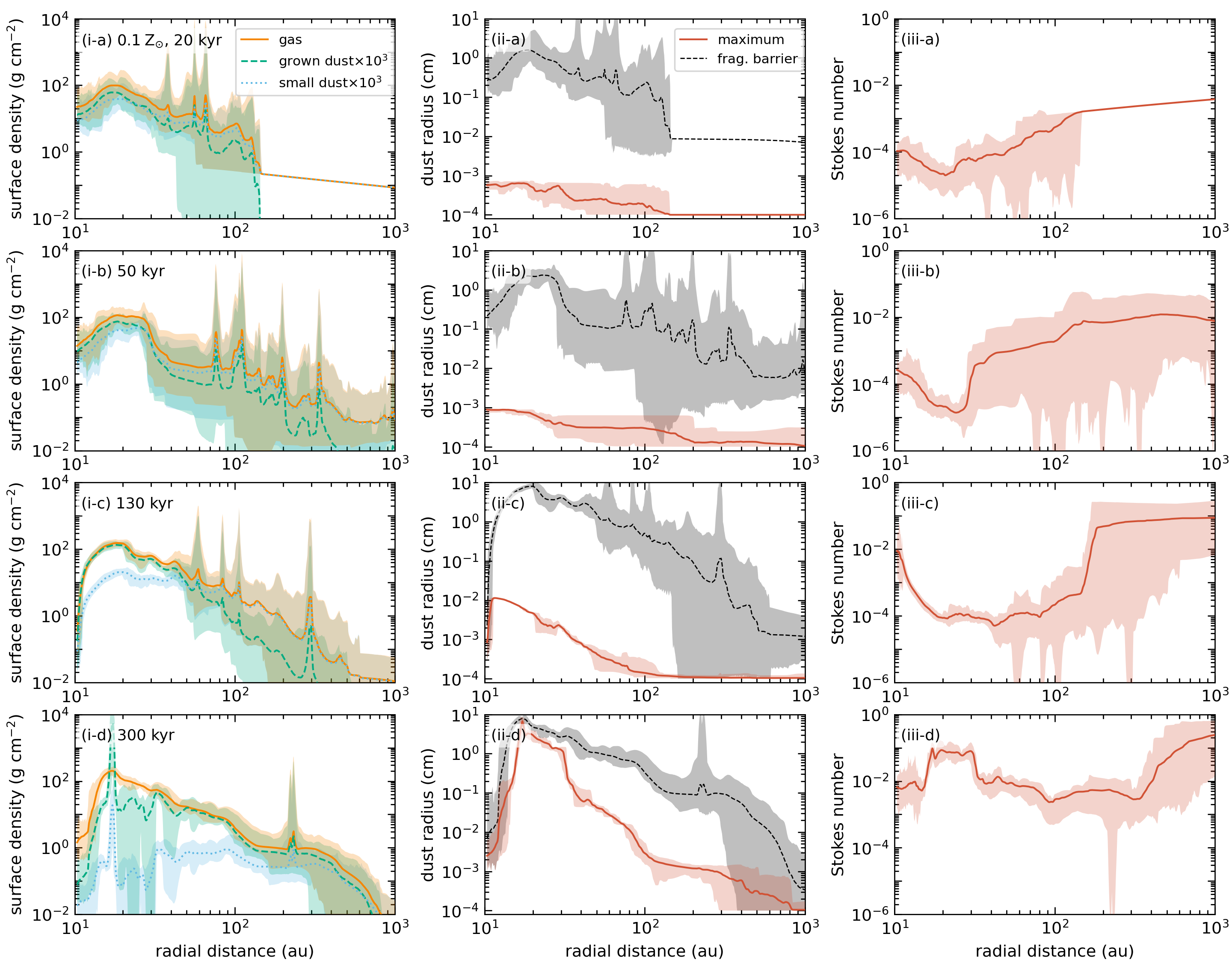}}
 \end{tabular}
 \caption{
 Same as Figure\:\ref{Fig:3} but for the case of 0.1\:$\Zsun$. The densities of the grown and small dust grains are multiplied by a factor of 10$^3$ to facilitate their comparison with the gas density.
 }
 \label{Fig:7}
 \end{center}
\end{figure*}

The left panels in Figure\:\ref{Fig:7} compares the surface densities of gas, grown dust, and small dust at the metallicity of 0.1\:$\Zsun$. As shown in panels i-a and i-b, at 0.1\:$\Zsun$, the surface density of the grown dust tends to be similar to or slightly lower than that of the small dust until 50\:kyr. By 130\:kyr (panel i-c), within a radius of 50\:au, the surface density of the grown dust exceeds that of the small dust. Later at 300\:kyr (panel i-d), the grown dust becomes the dominant component throughout the disk. Recall that, in the case of solar metallicity, the surface density of the grown dust is always greater than that of the small dust (Figure\:\ref{Fig:3}). These results indicate that at the metallicity of 0.1\:$\Zsun$, dust growth in the disk progresses at a slower rate.


Substituting the dust-to-gas mass ratio $\Sigma_{\rm d}/\Sigma_{\rm g} = 10^{-3}$ into \Eqref{Eq:t_growth}, we estimate the dust growth timescale at 0.1\:$\Zsun$ as $\sim$ 200\:kyr. In this study, we have followed such a long-term evolution over a period of 300\:kyr, comparable to this timescale. This marks the first time that such long-term simulations of dust growth in a low-metallicity disk have been performed alongside the formation and global evolution of the disk. Indeed, our simulation shows that the dust growth finally occurs after $\gtrsim$ 100\:kyr long-term evolution in a 0.1\:$\Zsun$ disk.


Panel i-d in Figure\:\ref{Fig:7} indicates that the surface density of grown dust has a spike at $\sim$15\:au, similar to the case with solar metallicity. This spike represents a ring-like structure, which is confirmed by the spatial distribution of grown dust. This dust ring is also formed by dust accumulation at a gas-pressure bump produced by the bottleneck effect. 


Despite slower dust growth at 0.1\:$\Zsun$ compared to the case with 1.0\:$\Zsun$, our simulation results suggest that the timing of the dust ring formation does not change. With solar metallicity, by 130\:kyr, dust grains sufficiently grow to a cm scale. Therefore, the first dust ring is formed at an early stage, $\sim$80\:kyr. However, since the disk is still gravitationally unstable at this period, clumps formed by gravitational instability orbit at radii outside the dust ring ($\sim$200\:au), destroying it. Subsequently, these clumps migrate inwards and are eventually dispersed by tidal forces from the central star. Dust grains from the dispersed clumps are distributed to the surrounding area, forming a new dust ring from these grains \citep{Vorobyov2019}. In the disk with solar metallicity, this cycle of dust ring formation and destruction is repeated, and a long-lasting dust ring is finally formed $\sim$300\:kyr, when the disk fragmentation diminishes. In the disk with 0.1\:$\Zsun$, since the dust growth timescale is similar to the timescale of the disk stabilization, the first dust ring is formed $\sim$230\:kyr and is maintained. This process results in density profiles with similar characteristics for both metallicities.


Slow dust growth at 0.1\:$\Zsun$ is also apparent in the middle panels in Figure\:\ref{Fig:7}. Panels ii-a and ii-b show that the maximum dust size is less than 10$^{-3}$\:cm in the inner disk at 20 and 50\:kyr. By 130\:kyr (panel ii-c), the size reaches approximately 10$^{-2}$\:cm near 10\:au but remains at 10$^{-4}$\:cm beyond 100\:au, indicating minimal dust growth in those regions. At 300\:kyr (panel ii-d), an overall dust growth is observed, with a maximum size of 10$^{-2}$ to 10$^{-3}$\:cm around 100\:au, 
regions where growth is limited in earlier epochs.


As shown in the right panels in Figure\:\ref{Fig:7}, the Stokes number in the early phase of the disk tends to be smaller than that at the solar metallicity, because of the slower dust growth. For example, at a position of 100\:au after 130\:kyr of disk formation (panel iii-c), the Stokes number is at 10$^{-4}$, three orders of magnitude smaller than that of the solar-metallicity case. 
At 300\:kyr when dust growth becomes significant (panel iii-d), the Stokes number in the disk region ($\sim$ 400\:au) increases, ranging between 10$^{-3}$ and 0.1. 

\begin{figure}
 \begin{center}
 \begin{tabular}{c} 
  {\includegraphics[width=0.94\columnwidth]{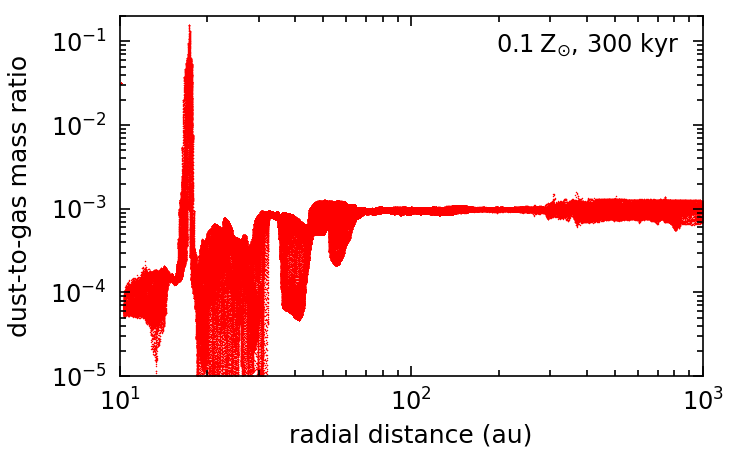}}
 \end{tabular}
 \caption{
 Same as Figure\:\ref{Fig:4} but for the case of 0.1\:$\Zsun$.
 }
 \label{Fig:8}
 \end{center}
\end{figure}

Finally, Figure\:\ref{Fig:8} shows that the dust-to-gas mass ratio in the dust ring is much greater than the canonical value of 0.001, increasing by two orders of magnitude to a maximum of 0.1. The magnitude of dust concentration is therefore comparable to that of the solar metallicity case, see Figure~\ref{Fig:4}, but the ring in this case is notably narrower. These high dust-to-gas mass ratios, together with the Stokes number exceeding 0.01, indicate that the ring may be prone to develop the streaming instability. On the other hand, the dust-to-gas ratio in the $0.1~Z_\odot$ disk beyond 70~au features little variations, unlike the 1.0~$Z_\odot$ case, indicating that dust dynamics in the lower metallicity is weakly perturbed by local pressure variations in the gravitationally unstable disk.

\subsubsection{One hundredth of solar metallicity}
\label{Sec:0.01Zsun}

\begin{figure}
 \begin{center}
 \begin{tabular}{c} 
  {\includegraphics[width=0.95\columnwidth]{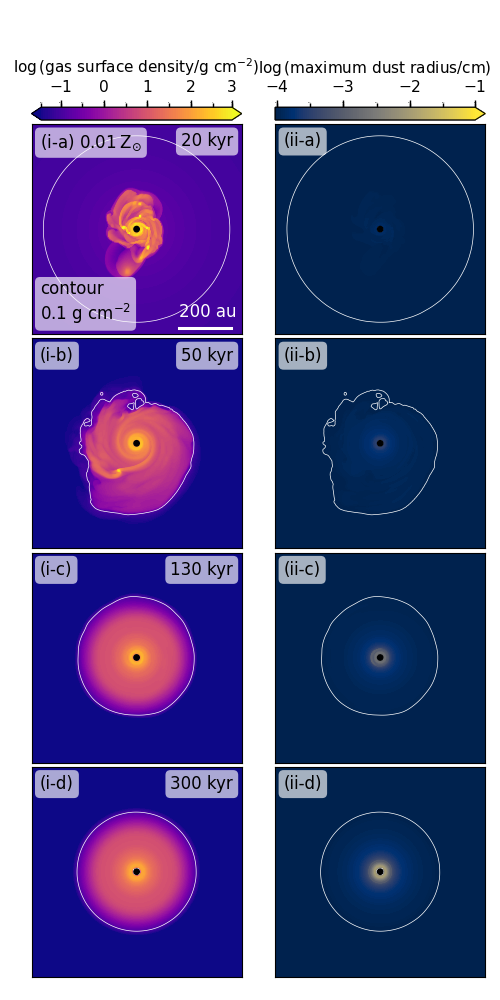}}
 \end{tabular}
 \caption{
 Same as Figure\:\ref{Fig:1} but for the case of 0.01\:$\Zsun$.
 }
 \label{Fig:9}
 \end{center}
\end{figure}


Figure\:\ref{Fig:9} demonstrates that the disk evolution at the metallicity of 0.01\:$\Zsun$ is distinctly different from the other two cases with 1.0 and 0.1\:$\Zsun$. In this model, the most intense fragmentation occurs at 20\:kyr (panel i-a). Afterward, the disk quickly becomes gravitationally stable, and at 130\:kyr (panel i-c), which represents the most unstable moment for the other models, the disk is completely stable, with both gas and dust having an axisymmetric surface density structure.


At 0.01\:$\Zsun$, the disk evolution proceeds more rapidly than the other models. This is attributed to higher gas temperatures in the envelope resulting from inefficient dust thermal emission induced by mutual collisions with gas molecules. The higher temperature of the envelope leads to rapid gas accretion from the envelope to the disk, which is approximately given by 
\begin{align}
\dot{M}_{\rm env} \simeq \frac{M_{\rm Jeans}}{t_{\rm ff}} \simeq \frac{c_{s}^3}{G} \propto T^{3/2},
\label{Eq:Mdot_env}
\end{align}
where $M_{\rm J}=4/3\pi\rho\left( l_{\rm Jeans}/2 \right)^3$ is the Jeans mass, $l_{\rm Jeans} = \sqrt{\pi c_{\rm s}^2 / \left( G\rho \right)}$ is the Jeans length, and $t_{\rm ff}=\sqrt{3\pi/\left( 32G\rho \right)}$ is the free-fall time \citep{Shu1977, Stahler1986}. Since the initial cloud mass for each model is fixed at $\simeq$ 0.9\:$\Msun$, under high accretion rates, the gas supply from the envelope to the disk concludes in an earlier phase, thereby leading to faster disk stabilization.


The panels in the right column of Figure\:\ref{Fig:9} indicate that at 0.01\:$\Zsun$ the dust size remains constant at 10$^{-4}$\:cm in both the disk and the surrounding envelope. By 130\:kyr (panel ii-c), only the innermost region of the disk ($\lesssim50$\:au) exhibits an increase in dust size to $\sim10^{-3}$\:cm. Throughout the evolution until 300\:kyr (panel ii-d), the region where dust size growth occurs does not expand, and the dust size only reaches 10$^{-2}$\:cm in the innermost part of the disk. Unlike the higher-metallicity cases, mm-sized dust grains are not observed throughout the disk at 0.01\:$\Zsun$.

\begin{figure}
 \begin{center}
 \begin{tabular}{c} 
  {\includegraphics[width=0.94\columnwidth]{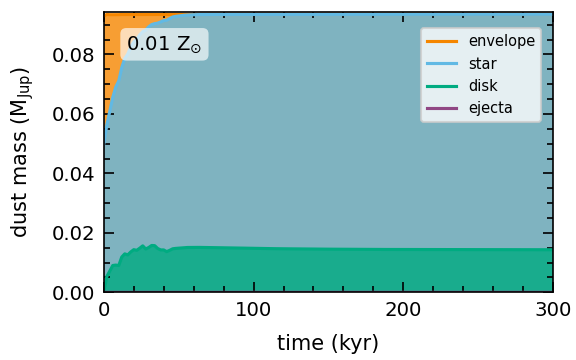}}
 \end{tabular}
 \caption{
 Same as Figure\:\ref{Fig:2} but for the cases of 0.01\:$\Zsun$.
 }
 \label{Fig:10}
 \end{center}
\end{figure}

We note that no ejection events are observed in the disk for this case. Figure\:\ref{Fig:10} indicates that all the dust grains either accrete onto the central star or remain within the disk. In the case of 0.01\:$\Zsun$, the shorter period of gravitational instability in the disk may contribute to making ejection events difficult.

\begin{figure*}
 \begin{center}
 \begin{tabular}{c} 
  {\includegraphics[width=1.98\columnwidth]{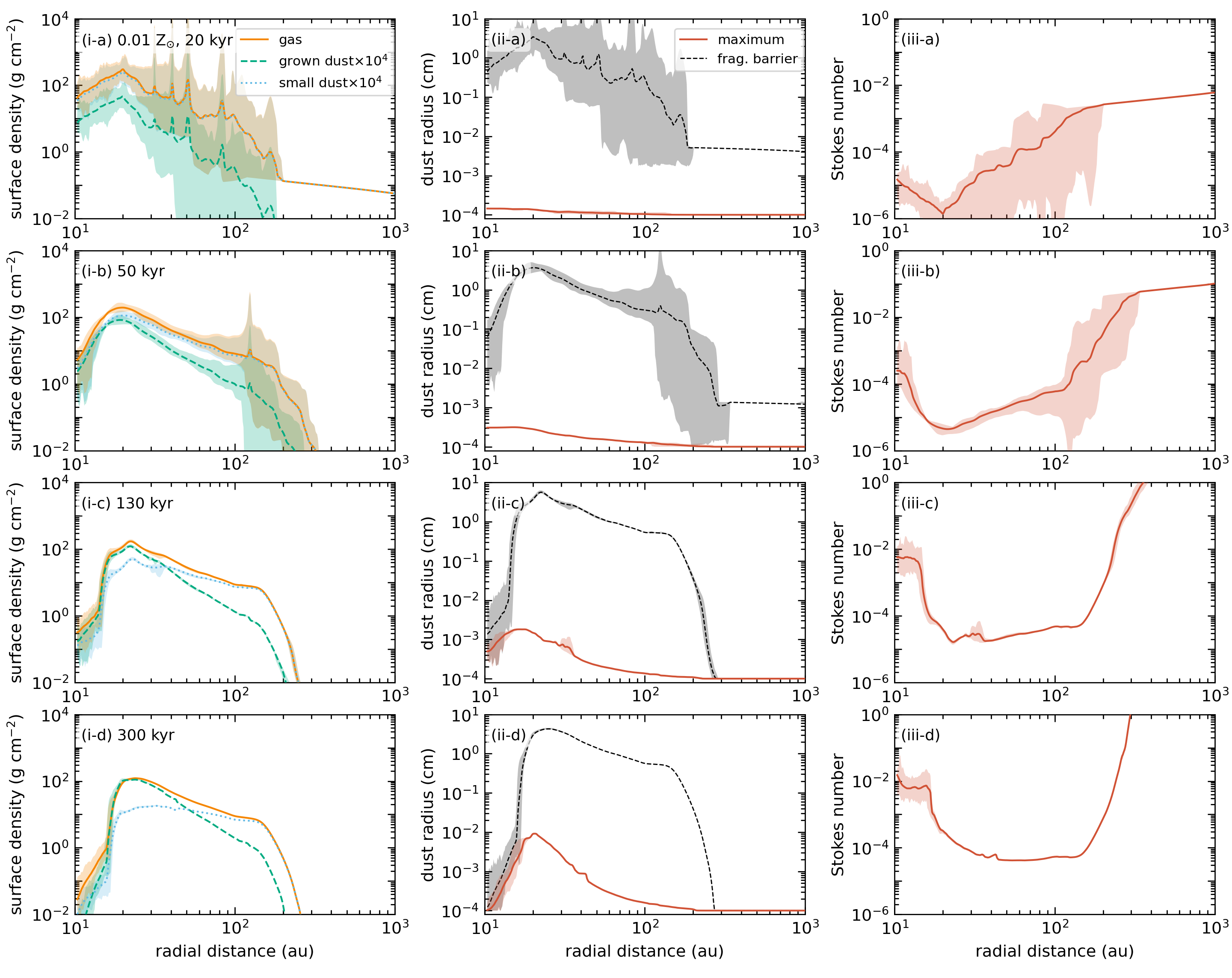}}
 \end{tabular}
 \caption{
 Same as Figure\:\ref{Fig:3} but for the case of 0.01\:$\Zsun$. 
 The densities of grown and small dust grains are multiplied by 10$^4$ to make the comparison with that of gas easier. 
 }
 \label{Fig:11}
 \end{center}
\end{figure*}

The left panels in Figure\:\ref{Fig:11} displays the surface-density radial profiles of gas, grown dust, and small dust at the metallicity of 0.01\:$\Zsun$. At 20\:kyr after disk formation (panel i-a), the surface density of the small dust is about an order of magnitude larger than that of the grown dust. Additionally, the surface density of the small dust multiplied by $10^4$ is nearly equivalent to the gas surface density, with the entire disk having a dust-to-gas mass ratio of 10$^{-4}$. At 50\:kyr (panel i-b), the surface density of the grown dust becomes similar to that of small dust within an inner radius of 20\:au. As time progresses to 130\:kyr and 300\:kyr (panels i-c and i-d), the grown dust becomes the dominant component within the radius of 50\:au. However, beyond this radius, the surface density of the grown dust is always smaller than that of the small dust and does not reach $10^{-4}$ of that of gas. 


The panels in the middle column of Figure\:\ref{Fig:11} indicate that the dust size does not grow much by the end of the simulation. According to panel ii-d, the maximum dust size in the innermost regions of the disk ($\sim$20\:au) is $\sim10^{-2}$\:cm, but it remains at $\sim10^{-4}$\:cm $\sim$100\:au. In contrast, we observe dust grains with mm scale in the cases of 1.0 and 0.1\:$\Zsun$. This clearly indicates slower dust growth at 0.01\:$\Zsun$ compared to the higher metallicity cases. These results are attributed to the simulated time of disk evolution (300\:kyr) being shorter than the dust growth timescale, which is $\sim$ 2\:Myr, with the dust-to-gas mass ratio of $10^{-4}$ (Equation\:\ref{Eq:t_growth}). The dust growth timescale at 0.01\:$\Zsun$ is comparable to the lifetime of protoplanetary disks in nearby clusters, $\sim$3--6\:Myr \citep[e.g.,][]{Haisch2001, Hernandez2007, Mamajek2009, Ribas2014}. Furthermore, shorter disk lifetimes in low-metallicity environments have been suggested by observations \citep[e.g.,][]{Yasui2010, Yasui2016, Yasui2021, Guarcello2021} and theories \citep[][]{Nakatani2018a, Nakatani2018b, Gehrig2023}. Therefore, the disk with the metallicity of 0.01\:$\Zsun$ may dissipate before dust grains can grow to mm in size. Panel iii-d in Figure\:\ref{Fig:11} presents that the Stokes number at 300\:kyr remains low at $10^{-4}$ across most of the disk ($\sim$ 20--100\:au) due to the small dust sizes. This value is two to three orders of magnitude smaller than those in the higher-metallicity cases.

\section{Implications for planetesimal and planet formation}
\label{Sec:planetesimal}


In the context of planet formation, how planetesimals are formed from mm- to cm-sized dust grains is an intriguing subject. The streaming instability is a promising pathway to planetesimal formation, as described in Section\:\ref{Sec:Intro}.


Several numerical studies examined dust clumping caused by streaming instability, utilizing the local shearing box approximation to model a small vertical slice of the protoplanetary disk \citep{Carrera2015, Yang2017, Li2021, Lim2023}. These studies explored a wide range of parameters, including the dust-to-gas mass ratio and the Stokes number, and identified the critical dust-to-gas mass ratio as a function of the Stokes number. Beyond this critical ratio, the streaming instability leads to strong clumping and subsequent dust self-gravitational collapse. The critical value is found to be lowest in the range of Stokes numbers between 0.1 and 0.3, with the dust-to-gas mass ratio ranging from 0.002 to 0.01. \citet{Lim2023} also considered the effect of turbulent gas diffusion on dust concentration caused by streaming instability and provided a fitting formula for the critical ratio.

\begin{figure}
 \begin{center}
 \begin{tabular}{c} 
  {\includegraphics[width=0.94\columnwidth]{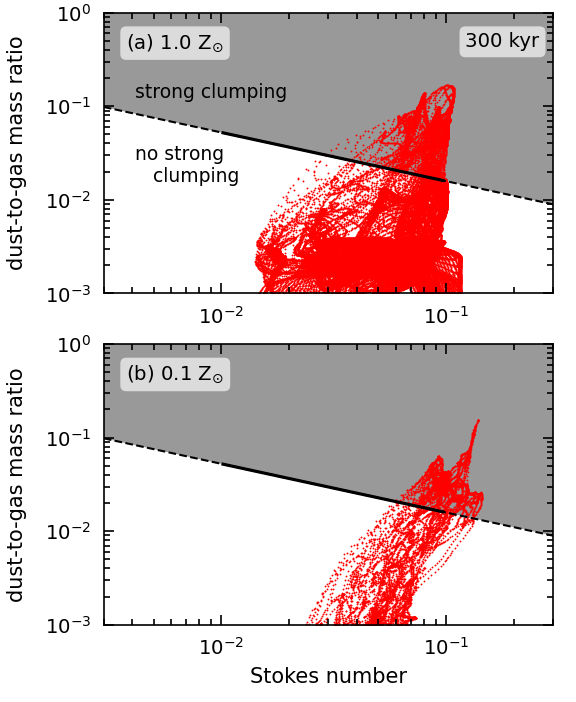}}
 \end{tabular}
 \caption{
 Possibility of planetesimal formation by streaming instability at 300\:kyr. The red points represent the grid cells from 10 to 40\:au from the center, including the dust ring regions. Each panel displays two different metallicities: 1.0\:$\Zsun$ (upper panel) and 0.1\:$\Zsun$ (lower panel). The black solid line depicts the critical dust-to-gas mass ratio as a function of the Stokes number provided by \citet{Lim2023}, above which dust concentration induced by streaming instability can trigger planetesimal formation. Since the range of the Stokes number explored in their study is from 0.01 to 0.1, we extrapolate the solid line and represent this extrapolation with the dashed line.
 }
 \label{Fig:12}
 \end{center}
\end{figure}


Dust rings with high surface densities facilitate dust growth and have a high dust-to-gas mass ratio, making them prone to the streaming instability and thus creating an ideal environment for planetesimal formation. To assess the potential for planetesimal formation in the dust rings observed in our simulations, we verify whether the grid cells within the dust ring regions meet the criterion identified by \citet{Lim2023}. \Figref{Fig:12} illustrates the distribution of grid cells from 10 to 40\:au from the center, including the dust ring regions, on the Stokes number -- dust-to-gas mass ratio plane at 300\:kyr. The gray shaded area indicates regions where the planetesimal formation through the streaming instability is expected. The figure reveals that some grid cells within this shaded area exist in the dust rings of both the metallicities. Therefore, we can anticipate planetesimal formation in both the dust rings. The total dust masses of grid cells that meet the critical dust-to-gas mass ratio are 68\:${\rm M}_\oplus$ for 1.0\:$\Zsun$ and 12\:${\rm M}_\oplus$ for 0.1\:$\Zsun$. The hydrodynamic simulations of streaming instability \citep{Simon2016, Abod2019} showed that the conversion efficiency from dust grains to planetesimals is approximately 50\%, with variations depending on the radial pressure gradient and ranging from 10 to 80\% \citep{Abod2019}. If a conversion efficiency of 50\% is assumed, then the total planetesimal masses in the dust ring would be 32\:${\rm M}_\oplus$ for 1\:$\Zsun$ and 6\:${\rm M}_\oplus$ for 0.1\:$\Zsun$.


\cite{Liu2019} examined protoplanet formation from planetesimals formed through streaming instability in a dust ring by conducting N-body simulations. In their simulations, planetesimals grow through mutual collisions with other planetesimals, i.e., planetesimal accretion, and through pebble accretion. The initial size distribution of planetesimals, formed by streaming instability, is described by a single power law \citep{Simon2016, Simon2017} and exhibits exponential decay at the high-mass end \citep{Schafer2017}. \cite{Liu2019} found that the largest planetesimal initially grows by planetesimal accretion, and then once its mass reaches 10$^{-2}$\:${\rm M}_\oplus$, pebble accretion becomes more efficient than planetesimal accretion. Eventually, a rocky protoplanet with a mass of 1\:${\rm M}_\oplus$ can form within 1\:Myr. They assumed the radial position of the dust ring is 2.7\:au, corresponding to the snow line for water ice. However, our dust rings are located far from the snow line. Since the efficiencies of planetesimal and pebble accretions decrease with radial distance (\citealt{Ormel2010} for protoplanet accretion; \citealt{Visser2016} for pebble accretion), protoplanet formation becomes more challenging. For example, the timescale for runaway planetesimal accretion is $\sim1$\:Myr at 20--30\:au, similar to the gas disk’s lifetime. This suggests that the gas disk may disperse before a growing planetesimal reaches 10$^{-2}$\:${\rm M}_\oplus$, preventing the switch to pebble accretion. Additionally, protoplanet formation in the disk with 0.1\:$\Zsun$ would be more difficult than in that with 1.0\:$\Zsun$, due to a reduced dust budget for pebble accretion. We note that a dust ring, formed by the bottleneck effect, appears at a radius where the efficiency of angular momentum transport is minimal. Consequently, its position may depend on the magnitude of alpha viscosity or the sink radius. A dust ring may form at smaller radii with other parameter values, and protoplanet formation may be allowed even in a low-metallicity disk.


As shown in \Figsref{Fig:9}{Fig:11}, in the case of metallicity 0.01\:$\Zsun$, dust growth is slow, and dust accumulation does not occur. Therefore, forming planetesimals through streaming instability is disadvantaged in the disk with 0.01\:$\Zsun$. Another mechanism for the planet formation is disk fragmentation caused by gravitational instability \citep{Boss1998, Rice2003, Mayer2007, Kratter2010, Machida2011, Zhu2012, Vorobyov2013, Tsukamoto2015, Nayakshin2017a, Stamatellos2018, Vorobyov2018}. Numerical and analytical studies examined the formation of gas giant planets through gravitational instability in a protoplanetary disk with low metallicity (\citealt{Boss2002, Matsuo2007, Meru2010, Vorobyov2020, Matsukoba2023} but see also \citealt{Cai2006}). In particular, \citet{Matsukoba2023} conducted numerical simulations of disk evolution over a long period (until 1\:Myr after disk formation) at a metallicity of 0.1\:$\Zsun$ and found a gas giant planet that survives until the end of simulations. In the results for metallicity 0.01\:$\Zsun$ in this study, all fragments fall into the central star, but depending on the initial conditions, some fragments may survive. For instance, increasing the strength of gas-cloud rotation to enlarge the disk's radius may make it less likely for fragments formed far from the center to fall into the central star.

\section{Summary}
\label{Sec:summary}

We have investigated dust growth in a protoplanetary disk by performing two-dimensional radiation-hydrodynamic simulations, following long-term evolution for 300\:kyr since the disk formation. Our code solves the dust-size growth and considers the dust motion separately from the gas, leading to non-uniform structures in the dust-to-gas mass ratio. Our simulations include three models with different initial dust-to-gas mass ratios: $10^{-2}$, $10^{-3}$, and $10^{-4}$. We specifically focus on the dependence of dust growth on these ratios. Our findings are summarized as follows:


\begin{enumerate}
   \item In the disk with the metallicity of 1.0\:$\Zsun$ (dust-to-gas mass ratio of $10^{-2}$), dust size growth is more efficient in the disk region than in the surrounding envelope due to the higher density and temperature, which lead to increased growth rates. The size of dust exceeds $10^{-2}$\:cm after 50\:kyr of disk formation at $< 200$\:au, consistent with the estimated timescale for dust size growth. The drift motion of grown dust grains results in a spatially non-uniform distribution of the dust-to-gas mass ratio. However, the deviation from the canonical values is not significant enough to affect the overall evolution of the disk.
   
   \item In the disk with the metallicity of 0.1\:$\Zsun$ (dust-to-gas mass ratio of $10^{-3}$), the growth rate of dust size is proportional to the dust density; hence, dust-size growth is slower compared to that at the metallicity of 1.0\:$\Zsun$. At this metallicity, the timescale for dust growth at the disk scale ($\sim$100\:au) is $\sim$200\:kyr. Once the disk age reaches this timescale, the dust size within the disk  becomes $>10^{-2}$\:cm, akin to distributions at the metallicity of 1.0\:$\Zsun$. The variation from the canonical dust-to-gas mass ratio is less pronounced than in the case of 1.0\:$\Zsun$, and similarly, the overall evolution of the disk remains unaffected by dust growth.
   
   \item In the disk with the metallicity of 0.01\:$\Zsun$ (dust-to-gas mass ratio of $10^{-4}$), dust size growth is observed only in the innermost regions ($\sim$20\:au). At a scale of 100\:au, the dust size remains $\sim10^{-4}$\:cm. The timescale for dust growth at this metallicity is $\sim$2\:Myr, which is comparable to the lifetime of the gas disk. Therefore, the gas disk may dissipate before the dust grains can grow to mm sizes.
   
   \item The dust ring forms in cases with metallicities of 1.0 and 0.1\:$\Zsun$. Within this ring, the dust density is enhanced, which facilitates dust growth, allowing dust size to exceed 1.0\:cm. Additionally, the Stokes number reaches 0.1. Such situations are conducive to streaming instability, planetesimals may be formed. 
\end{enumerate}


In this paper, we have discussed the growth of dust size across the disk. To perform high-resolution calculations at the 100\:au scale, we have introduced a sink cell of 10\:au at the center of the computational domain. Since dust size growth is more rapid in the inner regions, conducting calculations that resolve smaller radii allows us to observe dust size growth even in a disk with low metallicity. For instance, at a dust-to-gas mass ratio of $10^{-4}$, the timescale for dust size growth at a radius of 1\:au is estimated to be 10\:kyr. The growth of dust size on these smaller scales will be discussed in future studies.

\begin{acknowledgements}
This research could never be accomplished without the support by Grants-in-Aid for Scientific Research (TH: 19H01934, 19KK0353, 21H00041) from the Japan Society for the Promotion of Science. E.I.V. acknowledges support by the
Ministry of Science and Higher Education of the Russian
Federation (State assignment in the field of scientific activity 2023, GZ0110/23-10-IF). Numerical computations were carried out on Cray XC50 at the Center for Computational Astrophysics (CfCA) of the National Astronomical Observatory of Japan and the Vienna Scientific Cluster (VSC-4). 
\end{acknowledgements}


\bibliographystyle{aa} 
\bibliography{references}


\end{document}